\documentclass[english,onecolumn]{IEEEtran}
\usepackage[T1]{fontenc}
\usepackage[latin9]{inputenc}
\usepackage{amsmath}
\usepackage{amstext}
\usepackage{amssymb}
\usepackage{graphicx}
\usepackage{epsfig}
\usepackage{amsfonts}

\IEEEoverridecommandlockouts                              
\newtheorem{theorem}{Theorem}
\newtheorem{lemma}[theorem]{Lemma}
\newtheorem{notation}{Notation}

\title{\LARGE \bf Discontinuous integral control for mechanical systems}

\author{Jaime A. Moreno%
\thanks{J.~A. Moreno is with El\'{e}ctrica y Computaci\'{o}n, Instituto de Ingenier\'{\i}a, Universidad Nacional Aut\'{o}noma de M\'{e}xico, 04510 M\'{e}xico D.F., Mexico, JMorenoP@ii.unam.mx}
\thanks{A previous version of this paper has been submitted to the CDC2015.}
}

\begin{document}
\maketitle

\begin{abstract}
For mechanical systems we present a controller able to track an unknown
smooth signal, converging in finite time and by means of a continuous
control signal. The control scheme is insensitive against unknown
perturbations with bounded derivative. The controller consists of
a non locally Lipschitz state feedback control law, and a discontinuous
integral controller, that is able to estimate the unknown perturbation
and to compensate for it. To complete an output feedback control a
continuous observer for the velocity is added. It is shown that the
closed loop consisting of state feedback, state observer and discontinuous
integral controller has an equilibrium point that is globally, finite
time stable, despite of perturbations with bounded derivative. The
proof is based on a new smooth Lyapunov function. 
\end{abstract}

\section{Introduction}

We consider in this paper a second order system

\begin{equation}
\begin{array}{rcl}
\dot{\xi}_{1} & = & \xi_{2}\\
\dot{\xi}_{2} & = & f\left(\xi_{1},\,\xi_{2},\, t\right)+\rho\left(t\right)+\tau
\end{array}\label{ec:Plant}
\end{equation}
where $\xi_{1}\in\mathbb{R}$ and $\xi_{2}\in\mathbb{R}$ are the
states, $\tau\in\mathbb{R}$ is the control variable, $f\left(\xi_{1},\,\xi_{2},\, t\right)$
is some known function, while the term $\rho\left(t\right)$ corresponds
to uncertainties and/or perturbations. System \ref{ec:Plant} can
represent a mechanical system, where $\xi_{1}$ is the position and
$\xi_{2}$ is the velocity. An important control task is to track
a smooth time varying reference $r\left(t\right)$, i.e. if one defines
the tracking error $z_{1}=\xi_{1}-r$ and $z_{2}=\xi_{2}-\dot{r}$
the objective is to asymptotically stabilize the origin of system
\begin{equation}
\begin{array}{rcl}
\dot{z}_{1} & = & z_{2}\\
\dot{z}_{2} & = & f\left(\xi_{1},\,\xi_{2},\, t\right)+\rho\left(t\right)-\ddot{r}\left(t\right)+\tau\,.
\end{array}\label{ec:PlantTrack}
\end{equation}
With the control $\tau=u-f\left(\xi_{1},\,\xi_{2},\, t\right)+\ddot{r}\left(t\right)$
the system becomes 
\begin{equation}
\begin{array}{rcl}
\dot{x}_{1} & = & x_{2}\\
\dot{x}_{2} & = & u+\rho\left(t\right)\,,
\end{array}\label{ec:PlantFinal}
\end{equation}
where the perturbation $\rho\left(t\right)$ is a time varying signal,
not vanishing at the origin (i.e. when $x=0$ the perturbation can
still be acting). We notice that it is possible not to feed the second
derivative of the reference $\ddot{r}\left(t\right)$ to the control
$\tau$. In this case it will be considered as part of the perturbation
term $\rho\left(t\right)$.

Under the stated hypothesis it is well known that a continuous, memoryless
state feedback $u=k\left(x\right)$ is not able to stabilize $x=0$.
This is so, because the controller has to satisfy with the condition
$k\left(0\right)=0$, since the closed loop has to have an equilibrium
at the origin for vanishing perturbation. But if the perturbation
does not vanish, then the origin cannot be anymore an equilibrium
point. Discontinuous controllers, as the first order Sliding Mode
(SM) ones \cite{Utk92,Utk09} are able to solve the problem for non
vanishing (or persistently acting) bounded perturbations. However,
they require the design of a sliding surface that is reached in finite
time, but the target $x=0$ is attained only asymptotically fast,
and at the cost of a high frecuency switching of the control signal
(the so called \textit{chattering}), that has a negative effect in
the actuator, and excites unmodelled dynamics of the plant. Higher
Order Sliding Modes (HOSM) \cite{FriLev02,Lev03,Lev05,Lev07,Lev93}
provide a discontinuous controller for systems of relative degree
higher than one to robustly stabilize the origin $x=0$ despite of
bounded perturbations, but again at the expense of chattering. A natural
alternative consists in adding an integrator, i.e. defining a new
state $z=u+\rho\left(t\right)$, with $\dot{z}=v$ and designing a
third order HOSM controller for the new control variable $v$. This
allows to reach the origin in finite time, and it will be insensitive
to Lipschitz perturbations, i.e. with $\dot{\rho}\left(t\right)$
bounded. In this form a continuous control signal $u$ will be obtained,
so that the chattering effect is reduced. However, this requires feedback
not only the two states $x_{1}$ and $x_{2}$ but also the state $z$,
which is unknown due to the unknown perturbation. Moreover, to implement
an output feedback controller (assuming that only the position $x_{1}$
is measured) it is necessary to differentiate two times the position
$x_{1}$, with the consequent noise amplification effect.

In the case of (almost) constant perturbations $\rho\left(t\right)$
a classical solution to the robust regulation problem is the use of
integral action, as for example in the PID control\cite{Kha02}. The
linear solution would consist of a state feedback plus an integral
action, $u=-k_{1}x_{1}-k_{2}x_{2}+z\,,\,\dot{z}=-k_{3}x_{1}$. This
controller requires only to feedback the position and the velocity.
For an output feedback it would be only necessary to estimate the
velocity (with the D action for example). In contrast to the HOSM
controller this PID control is only able to reject constant perturbations,
instead of Lipschitz ones, and it will reach the target only exponentially
fast, and not in finite time. By the Internal Model Principle it would
be possible to reject exactly any kind of time varying perturbations
$\rho\left(t\right)$, for which a dynamical model (so called an exosystem)
is available. However this would increase the complexity (order) of
the controller, since this exosystem has to be included in the control
law.

Here we provide a solution to the problem, that is somehow an intermediate
solution between HOSM and PID control. Similar to the HOSM control
our solution uses a discontinuous integral action, it can compensate
perturbations with bounded derivative ($\rho\left(t\right)$ is Lipschitz)
and the origin is reached in finite time. So it can solve not only
regulation problems (where $\rho$ is constant) but also tracking
problems (with $\rho$ time varying) in finite time and with the same
complexity of the controller. Similar to the PID control the proposed
controller provides a continuous control signal (avoiding chattering)
and it requires only to feedback position and velocity. We also provide
for a (non classical) D term, i.e. a finite time converging observer,
to estimate the velocity. This basic idea has been already presented
in our previous work \cite{ZamMor13}. In the present one we give
a much simpler Lyapunov-based proof, and we also include an observer
in the closed loop together with its Lyapunov proof. Our solution
can be seen as a generalization of the Super Twisting control for
systems of relative degree one \cite{Lev93,FriLev02,Lev03,Lev07,MorOso12}
to systems with relative degree two.

The rest of the paper is organized as follows. In next Section \ref{sec:DiscontinuousIC}
we present the main result: The Discontinuous Integral Controller
with and without observer and give some discussion on the algorithms.
Section \ref{sec:Proof} is dedicated to present the Lyapunov-based
proof of the convergence of the closed loop for the proposed control
algorithms. In Section \ref{Sec:sim} we give an illustrative example
with some simulations and in Section \ref{sec:Conclusion} some conclusions
are drawn.

\begin{notation} To simplify the presentation we introduce the following
notation. For a real variable $z\in\mathbb{R}$ and a real number
$p\in\mathbb{R}$ the symbol $\lfloor z\rceil^{p}=|z|^{p}sign(z)$,
i.e. the signed power $p$ of $z$. Note that $\lfloor z\rceil^{2}=|z|^{2}sgn(z)\neq z^{2}$,
and if $p$ is an odd number then $\lfloor z\rceil^{p}=z^{p}$. Note
also in particular, that 
\[
\begin{array}{rclrcl}
\lfloor z\rceil^{0} & = & sgn(z), & \lfloor z\rceil^{0}z^{p} & = & |z|^{p}\\
\lfloor z\rceil^{0}|z|^{p} & = & \lfloor z\rceil^{p}, & \lfloor z\rceil^{p}\lfloor z\rceil^{q} & = & |z|^{p+q}
\end{array}\,.
\]
\end{notation}

\section{Discontinuous Integral Controller}

\label{sec:DiscontinuousIC}For the robust finite time stabilization
of the origin of system (\ref{ec:PlantFinal}) we propose a nonlinear,
homogeneous state feedback control law, which is able to stabilize
the origin in finite time in the absence of non vanishing perturbations,
and a discontinuous integral controller is added to compensate for
the persistently acting perturbations. In contrast to the continuous
integral controller, that can only compensate for (almost) constant
perturbations, the discontinuous one can deal with time varying perturbations
which are Lipschitz continuous, that is, their derivatives exist almost
everywhere and it is uniformly bounded. The control signal of the
controller is continuous, so that the chattering effect of the SM
and HOSM controllers is avoided. 
\begin{theorem}
\label{th:CID} Consider the plant (\ref{ec:PlantFinal}) with Lipschitz
continuous perturbation signal $\rho\left(t\right)$ with Lispchitz
constant $L$. Then the control law 
\begin{equation}
\begin{array}{rcl}
u & = & -k_{1}\left\lceil x_{1}\right\rfloor ^{\frac{1}{3}}-k_{2}\left\lceil x_{2}\right\rfloor ^{\frac{1}{2}}+z\\
\dot{z} & = & -k_{3}\left\lceil x_{1}+k_{4}x_{2}^{\frac{3}{2}}\right\rfloor ^{0}
\end{array}\label{ec:CID}
\end{equation}
can stabilize the origin in finite time for any $k_{4}$ and appropriate
designed gains $k_{1}$, $k_{2}$, $k_{3}$. \hfill  $\blacksquare$
\end{theorem}

This Theorem shows that with the addition of the discontinuous integral
term it is possible to eliminate completely the effect of the Lipschitz
perturbation $\delta\left(t\right)$, that cannot be (fully) compensated
by the state feedback $u=-k_{1}\left\lceil x_{1}\right\rfloor ^{\frac{1}{3}}-k_{2}\left\lceil x_{2}\right\rfloor ^{\frac{1}{2}}$
alone. In fact, the integral controller can be interpreted as a perturbation
estimator, since $z\left(t\right)=-\delta\left(t\right)$ after a
finite time. 

It is remarkable, that the observer itself has not been designed to
be robust against the perturbation, but the discontinuous integral
controller is able to compensate for it. Note also that the input
to the discontinuous integrator $y=x_{1}+k_{4}x_{2}^{\frac{3}{2}}$
can be a combination of the position (with relative degree two) and
the velocity (with a relative degree one). The value of $k_{4}$ can
be arbitrary (including zero), so that the velocity is not necessary
for the integral action. For $k_{4}>0$ this output can be seen as
a passive output of the system $\left(x_{1},\, x_{2}\right)$. However,
it is necessary to have the position in this signal, otherwise the
closed loop will be unstable. 

By performing a linear transformation of the plant (\ref{ec:PlantFinal})
with controller (\ref{ec:CID}) $\xi=\lambda x$, for any some $\lambda>0$,
it is easy to show that if the gains $\left(k_{1},\, k_{2},\, k_{3},\, k_{4}\right)$
achieve the objective for a perturbation with Lipschitz constant $L$,
then the gains $\left(\lambda^{\frac{2}{3}}k_{1},\,\lambda^{\frac{1}{2}}k_{2},\,\lambda k_{3},\,\lambda^{-\frac{3}{2}}k_{4}\right)$
will also stabilize the system for a perturbation with Lipschitz constant
$\lambda L$.

The implementation of controller (\ref{ec:CID}) requires the mesurement
of both states $x_{1},\, x_{2}$. If only the position is measured
a finite time convergent observer for $x_{2}$ can be implemented,
so that an output feedback control is obtained.
\begin{theorem}
\label{thm:OFC}Consider the plant (\ref{ec:PlantFinal}) with Lipschitz
continuous perturbation signal $\rho\left(t\right)$ with Lispchitz
constant $L$. Then the output feedback control law 
\begin{equation}
\begin{array}{rcl}
\dot{\hat{x}}_{1} & = & -l_{1}\left\lceil \hat{x}_{1}-x_{1}\right\rfloor ^{\frac{2}{3}}+\hat{x}_{2}\\
\dot{\hat{x}}_{2} & = & -l_{2}\left\lceil \hat{x}_{1}-x_{1}\right\rfloor ^{\frac{1}{3}}-k_{1}\left\lceil x_{1}\right\rfloor ^{\frac{1}{3}}-k_{2}\left\lceil \hat{x}_{2}\right\rfloor ^{\frac{1}{2}}\\
u & = & -k_{1}\left\lceil x_{1}\right\rfloor ^{\frac{1}{3}}-k_{2}\left\lceil \hat{x}_{2}\right\rfloor ^{\frac{1}{2}}+z\\
\dot{z} & = & -k_{3}\left\lceil x_{1}+k_{4}\hat{x}_{2}^{\frac{3}{2}}\right\rfloor ^{0}\,,
\end{array}\label{ec:CID-Obs}
\end{equation}
can stabilize the origin in finite time for appropriate designed gains
$k_{1}$, $k_{2}$, $k_{3}$, $k_{4}$, $l_{1}$ and $l_{2}$. \hfill  $\blacksquare$
\end{theorem}
Similarly to the previous case it follows that if the gains $\left(k_{1},\, k_{2},\, k_{3},\, k_{4},\, l_{1},\, l_{2}\right)$
achieve the objective for a perturbation with Lipschitz constant $L$,
then the gains $\left(\lambda^{\frac{2}{3}}k_{1},\,\lambda^{\frac{1}{2}}k_{2},\,\lambda k_{3},\,\lambda^{-\frac{3}{2}}k_{4},\,\lambda^{\frac{1}{3}}l_{1},\,\lambda^{\frac{2}{3}}l_{2}\right)$
will also stabilize the system for a perturbation with Lipschitz constant
$\lambda L$. System (\ref{ec:PlantFinal}) with controller (\ref{ec:CID-Obs})
is given by the dynamics

\begin{equation}
\begin{array}{rcl}
\dot{x}_{1} & = & x_{2}\\
\dot{x}_{2} & = & -k_{1}\left\lceil x_{1}\right\rfloor ^{\frac{1}{3}}-k_{2}\left\lceil \hat{x}_{2}\right\rfloor ^{\frac{1}{2}}+z+\rho\left(t\right)\,,\\
\dot{\hat{x}}_{1} & = & -l_{1}\left\lceil \hat{x}_{1}-x_{1}\right\rfloor ^{\frac{2}{3}}+\hat{x}_{2}\\
\dot{\hat{x}}_{2} & = & -l_{2}\left\lceil \hat{x}_{1}-x_{1}\right\rfloor ^{\frac{1}{3}}-k_{1}\left\lceil x_{1}\right\rfloor ^{\frac{1}{3}}-k_{2}\left\lceil \hat{x}_{2}\right\rfloor ^{\frac{1}{2}}\\
\dot{z} & = & -k_{3}\left\lceil x_{1}+k_{4}\hat{x}_{2}^{\frac{3}{2}}\right\rfloor ^{0}
\end{array}\label{ec:Plant+CID-Obs}
\end{equation}
which is a discontinuous system, whose trajectories are defined in
the sense of Filippov \cite{Fil98}. If we introduce the estimation
errors $e_{1}=\hat{x}_{1}-x_{1}$, $e_{2}=\hat{x}_{2}-x_{2}$ and
the effect of the perturbation $x_{3}=z+\rho\left(t\right)$, then
the dynamics of the system can be rewritten as
\begin{equation}
\begin{array}{rcl}
\dot{x}_{1} & = & x_{2}\\
\dot{x}_{2} & = & -k_{1}\left\lceil x_{1}\right\rfloor ^{\frac{1}{3}}-k_{2}\left\lceil \hat{x}_{2}\right\rfloor ^{\frac{1}{2}}+x_{3}\,,\\
\dot{e}_{1} & = & -l_{1}\left\lceil e_{1}\right\rfloor ^{\frac{2}{3}}+e_{2}\\
\dot{e}_{2} & = & -l_{2}\left\lceil e_{1}\right\rfloor ^{\frac{1}{3}}-x_{3}\left(t\right)\\
\dot{x}_{3} & = & -k_{3}\left\lceil x_{1}+k_{4}\left(x_{2}+e_{2}\right)^{\frac{3}{2}}\right\rfloor ^{0}+\dot{\rho}\left(t\right)
\end{array}\label{ec:CLError}
\end{equation}
We will prove Theorem \ref{thm:OFC} by showing that $\left(x_{1},\, x_{2},\, e_{1},\, e_{2},\, x_{3}\right)=0$,
which is an equilibrium point, is Globally Finite Time Stable. An
important property of the previous systems is the homogeneity, that
we recall briefly. 

For a given vector $x=(x_{1},...,x_{n})^{T}\in\mathbb{R}^{n}$, the
dilation operator is defined as $\Delta_{\epsilon}^{r}x:=(\epsilon^{r_{1}}x_{1},...,\epsilon^{r_{n}}x_{n})$,
$\forall\epsilon>0$, where $r_{i}>0$ are the weights of the coordinates.
Let $\textbf{r}=(r_{1},...,r_{n})$ be the vector of weights. A function
$V:\mathbb{R}^{n}\rightarrow\mathbb{R}$ (respectively, a vector field
$f:\mathbb{R}^{n}\rightarrow\mathbb{R}^{n}$, or a vector-set field
$F(x)\subset\mathbb{R}^{n}$) is called $\textbf{r}$-homogeneous
of degree $m\in\mathbb{R}$ if the identity $V(\Delta_{\epsilon}^{r}x)=\epsilon^{m}V(x)$
holds (resp., $f(\Delta_{\epsilon}^{r}x)=\epsilon^{l}\Delta_{\epsilon}^{r}f(x)$,
or $F(\Delta_{\epsilon}^{r}x)=\epsilon^{l}\Delta_{\epsilon}^{r}F(x)$),
\cite{BacRos05,Lev05}. Consider that the vector $\textbf{r}$ and
dilation $\Delta_{\epsilon}^{r}x$ are fixed. The homogeneous norm
is defined by $\left\Vert x\right\Vert _{\mathbf{r},\, p}:=\left(\sum_{i=1}^{n}\left|x_{i}\right|^{\frac{p}{r_{i}}}\right)^{\frac{1}{p}},\,\forall x\in\mathbb{R}^{n}$,
for any $p\geq1$. The set $S=\{x\in\mathbb{R}^{n}:\left\Vert x\right\Vert _{\mathbf{r},\, p}=1\}$
is the corresponding unit sphere. Homogeneous systems have important
properties as e.g. that local stability implies global stability and
if the homogeneous degree is negative asymptotic stability implies
finite time stability \cite{BacRos05,Lev05}: Assume that the origin
of a Filippov Differential Inclusion, $\dot{x}\in F(x)$, is strongly
locally Asymptotic Stable and the vector-set field $F$ is \textbf{$r$}-homogeneous
of degree $l<0$; then, $x=0$ is strongly globally finite-time stable
and the settling time is continuous at zero and locally bounded.

System (\ref{ec:CLError}) is homogeneous, with weights $\textbf{r}=(3,\,2,\,3,\,2,\,1)$
for the variables $\left(x_{1},\, x_{2},\, e_{1},\, e_{2},\, x_{3}\right)$
and negative homogeneous degree $l=-1$. From homogeneity arguments
\cite{Lev05,FriLev02,Lev03} one expects that the controllers have
precision of order $\left|x_{1}\right|\leq\nu_{1}\tau^{3}$ and $\left|x_{2}\right|\leq\nu_{2}\tau^{2}$,
where $\tau$ is the discretization step and $\nu_{1}$ and $\nu_{2}$
are constants depending only on the gains of the algorithm. Moreover,
it is easy to show that for the Lyapunov functions the following inequality
is satisfied
\[
{\displaystyle \dot{V}(x)\leq-\kappa V^{\frac{4}{5}}(x)}\,,
\]
from which finite time convergence can be deduced. With the value
of $\kappa$ it is possible to estimate the convergence time as
\[
T(x_{0})\leq\frac{5}{\kappa}V^{\frac{1}{5}}(x_{0})\,.
\]

\section{Lyapunov function for the closed loop system}

\label{sec:Proof}We show, by using homogeneous and smooth Lyapunov
Functions, that Theorems \ref{th:CID} and \ref{thm:OFC} are valid.

\subsection{Proof of Theorem \ref{th:CID}}

Consider the closed loop system of plant (\ref{ec:PlantFinal}) with
the controller (\ref{ec:CID}), with the variable $x_{3}=z+\rho\left(t\right)$
\begin{equation}
\begin{array}{rcl}
\dot{x}_{1} & = & x_{2}\\
\dot{x}_{2} & = & -k_{1}\left\lceil x_{1}\right\rfloor ^{\frac{1}{3}}-k_{2}\left\lceil x_{2}\right\rfloor ^{\frac{1}{2}}+x_{3}\\
\dot{x}_{3} & = & -k_{3}\left\lceil x_{1}+k_{4}x_{2}^{\frac{3}{2}}\right\rfloor ^{0}+\dot{\rho}\left(t\right)\,.
\end{array}\label{ec:CLError-Thm1}
\end{equation}
Consider the homogeneous and smooth Lyapunov Function
\begin{eqnarray*}
V\left(x_{1},\, x_{2},\, x_{3}\right) & = & \gamma_{1}\left|\xi_{1}\right|^{\frac{5}{3}}+\gamma_{12}\xi_{1}x_{2}+\left|x_{2}\right|^{\frac{5}{2}}+\frac{1}{5}\left|x_{3}\right|^{5}\,,
\end{eqnarray*}
where $\xi_{1}=x_{1}-\frac{1}{k_{1}^{3}}\left\lceil x_{3}\right\rfloor ^{3}$.
We recall Young's inequality. 
\begin{lemma}
\cite{HarLit51} \label{lem:Young}For any positive real numbers $a>0$,
$b>0$, $c>0$, $p>1$ and $q>1$, with $\frac{1}{p}+\frac{1}{q}=1$,
the following inequality is always satisfied 
\[
ab\leq c^{p}{\displaystyle \frac{a^{p}}{p}+c^{-q}{\displaystyle \frac{b^{q}}{q}\,.}} \qquad \qquad \blacksquare
\] 

\end{lemma}
From Lemma \ref{lem:Young} it follows easily that given $\gamma_{12}$
it is always possible to render $V$ positive definite selecting $\gamma_{1}$
sufficiently large.

Its derivative along the trajectories of (\ref{ec:CLError-Thm1})
is given by
\begin{eqnarray*}
\dot{V} & = & W_{1}\left(\xi_{1},\, x_{2}\right)+W_{2}\left(\xi_{1},\, x_{2},\, x_{3}\right)+W_{3}\left(x,\,\dot{\rho}\right)\,,
\end{eqnarray*}
where
\begin{eqnarray*}
W_{1} & = & \left(\frac{5}{3}\gamma_{1}\left\lceil \xi_{1}\right\rfloor ^{\frac{2}{3}}+\gamma_{12}x_{2}\right)x_{2} 
-\frac{5}{2}k_{2}\left(\frac{2}{5}\gamma_{12}\xi_{1}+\left\lceil x_{2}\right\rfloor ^{\frac{3}{2}}\right)\left(\frac{k_{1}}{k_{2}}\left\lceil \xi_{1}\right\rfloor ^{\frac{1}{3}}+\left\lceil x_{2}\right\rfloor ^{\frac{1}{2}}\right)
\end{eqnarray*}
\begin{eqnarray*}
W_{2} & = & -k_{1}\left(\gamma_{12}\xi_{1}+\frac{5}{2}\left\lceil x_{2}\right\rfloor ^{\frac{3}{2}}\right)\alpha\left(\xi_{1},\, x_{3}\right)\,,
\end{eqnarray*}
\[
\alpha\left(\xi_{1},\, x_{3}\right)=\left\lceil \xi_{1}+\frac{1}{k_{1}^{3}}\left\lceil x_{3}\right\rfloor ^{3}\right\rfloor ^{\frac{1}{3}}-\left\lceil \frac{1}{k_{1}^{3}}\left\lceil x_{3}\right\rfloor ^{3}\right\rfloor ^{\frac{1}{3}}-\left\lceil \xi_{1}\right\rfloor ^{\frac{1}{3}}\,,
\]
\begin{eqnarray*}
W_{3}\left(x,\,\dot{\rho}\right) & = & \left(k_{3}\left\lceil x_{1}+k_{4}x_{2}^{\frac{3}{2}}\right\rfloor ^{0}-\dot{\rho}\left(t\right)\right)\left|x_{3}\right|^{2} 
 \left[\frac{3}{k_{1}^{3}}\left(\frac{5}{3}\gamma_{1}\left\lceil \xi_{1}\right\rfloor ^{\frac{2}{3}}+\gamma_{12}x_{2}\right)-\left\lceil x_{3}\right\rfloor ^{2}\right]\,.
\end{eqnarray*}

Consider first $W_{1}$. If we set 
\[
\gamma_{12}=\frac{5}{2}\left(\frac{k_{1}}{k_{2}}\right)^{3}
\]
then we get
\begin{eqnarray*}
W_{1} & = & \frac{5}{2}\left(\frac{k_{1}}{k_{2}}\right)^{3}\left(\frac{2}{3}\left(\frac{k_{2}}{k_{1}}\right)^{3}\gamma_{1}\left\lceil \xi_{1}\right\rfloor ^{\frac{2}{3}}+x_{2}\right)x_{2}-\\
 &  & \frac{5}{2}k_{2}\left(\left(\frac{k_{1}}{k_{2}}\right)^{3}\xi_{1}+\left\lceil x_{2}\right\rfloor ^{\frac{3}{2}}\right)\left(\frac{k_{1}}{k_{2}}\left\lceil \xi_{1}\right\rfloor ^{\frac{1}{3}}+\left\lceil x_{2}\right\rfloor ^{\frac{1}{2}}\right)
\end{eqnarray*}
The second term is negative semidefinite in $\left(\xi_{1},\, x_{2}\right)$,
and it becomes zero only at the set ${\cal S}_{1}=\left\{ x_{2}=-\left(\frac{k_{1}}{k_{2}}\right)^{2}\left\lceil x_{1}-\frac{1}{k_{1}^{3}}\left\lceil x_{3}\right\rfloor ^{3}\right\rfloor ^{\frac{2}{3}}\right\} $.
On ${\cal S}_{1}$ the value of $W_{1}$ becomes
\begin{eqnarray*}
W_{1}\mid_{{\cal S}_{1}} & = & -\frac{5}{2}\left(\frac{k_{1}}{k_{2}}\right)^{5}\left(\frac{2}{3}\left(\frac{k_{2}}{k_{1}}\right)^{3}\gamma_{1}-\left(\frac{k_{1}}{k_{2}}\right)^{2}\right)\left|\xi_{1}\right|^{\frac{4}{3}}\,,
\end{eqnarray*}
which is negative if
\[
\gamma_{1}>\frac{3}{2}\left(\frac{k_{1}}{k_{2}}\right)^{5}\,,
\]
i.e. $\gamma_{1}$ is set sufficiently large. We recall the following
well-known property of homogeneous functions
\begin{lemma}
\label{lemAndrieuH} \cite{Hes66,AndPra08} Let $\eta:\mathbb{R}^{n}\rightarrow\mathbb{R}$
and $\gamma:\mathbb{R}^{n}\rightarrow\mathbb{R}_{+}$ be two homogeneous
functions, with weights $\textbf{r}=(r_{1},...,r_{n})$ and degrees
$m$, such that the following holds 
\[
\begin{array}{c}
\{x\in\mathbb{R}^{n}\setminus\{0\}:\gamma(x)=0\}\subseteq\{x\in\mathbb{R}^{n}\setminus\{0\}:\eta(x)<0\},\end{array}
\]
Then, there exists a real number $\lambda^{\ast}$ such that, for
all $\lambda\leq\lambda^{\ast}$ for all $x\in\mathbb{R}^{n}\setminus\{0\}$,
and some $c>0$, $\eta(x)-\lambda\gamma(x)<-c\left\Vert x\right\Vert _{\mathbf{r},\, p}^{m}$.  \hfill  $\blacksquare$
\end{lemma}

Using Lemma \ref{lemAndrieuH} it follows that $W_{1}<-c\left\Vert \left(\xi_{1},\, x_{2}\right)\right\Vert _{\mathbf{r},\, p}^{4}$
for $k_{2}$ sufficiently large.

Now we consider the continuous function, appearing in the third term
\[
\alpha\left(\xi_{1},\, x_{3}\right)=\left\lceil \xi_{1}+\frac{1}{k_{1}^{3}}\left\lceil x_{3}\right\rfloor ^{3}\right\rfloor ^{\frac{1}{3}}-\left\lceil \frac{1}{k_{1}^{3}}\left\lceil x_{3}\right\rfloor ^{3}\right\rfloor ^{\frac{1}{3}}-\left\lceil \xi_{1}\right\rfloor ^{\frac{1}{3}}\,,
\]
which is homogeneous of degree $1$. Note that $\alpha\left(\xi_{1},\,0\right)=0$
and also $\alpha\left(0,\, x_{3}\right)=0$. Since $\alpha\left(\xi_{1},\, x_{3}\right)$
is homogeneous in $\xi_{1}$ it follows that
\[
\left|\alpha\left(\xi_{1},\, x_{3}\right)\right|\leq\delta\left(x_{3}\right)\left|\xi_{1}\right|^{\frac{1}{3}}\,,
\]
where $\delta\left(x_{3}\right)\geq0$ is continuous in $x_{3}$ and
$\delta\left(0\right)=0$. This implies that, for some $\beta>0$
\[
\left|W_{2}\right|\leq\delta\left(x_{3}\right)\beta\left\Vert \left(\xi_{1},\, x_{2}\right)\right\Vert _{\mathbf{r},\, p}^{4}\,,
\]
and therefore, for small values of $x_{3}$
\begin{eqnarray*}
W_{1}+W_{2} & \leq & -\left(c-\beta\delta\left(x_{3}\right)\right)\left\Vert \left(\xi_{1},\, x_{2}\right)\right\Vert _{\mathbf{r},\, p}^{4}\\
 & \leq & -c_{2}\left\Vert \left(\xi_{1},\, x_{2}\right)\right\Vert _{\mathbf{r},\, p}^{4}\,.
\end{eqnarray*}
Due to the homogeneity of $W_{1}+W_{2}$ this must be valid globally.
Note that $W_{1}\left(\xi_{1},\, x_{2}\right)+W_{2}\left(\xi_{1},\, x_{2},\, x_{3}\right)=0$
on the set ${\cal S}_{2}=\left\{ \left(\xi_{1},\, x_{2}\right)=0\right\} $.
Finally, the value of $W_{3}$ on the set ${\cal S}_{2}$ is given
by
\begin{eqnarray*}
W_{3}\mid_{{\cal S}_{2}} & \leq & -\left(k_{3}-L\right)\left|x_{3}\right|^{4}\,,
\end{eqnarray*}
which is negative for $L<k_{3}$. Again, Lemma \ref{lemAndrieuH}
implies that $\dot{V}<0$ selecting $k_{3}$ (and $L$) sufficiently
small.

\subsection{Proof of Theorem \ref{thm:OFC}}

We first prove that the closed loop system, without perturbation and
without integral controller, that we can write as (see (\ref{ec:CLError}))
\begin{equation}
\begin{array}{rcl}
\dot{x}_{1} & = & x_{2}\\
\dot{x}_{2} & = & -k_{1}\left\lceil x_{1}\right\rfloor ^{\frac{1}{3}}-k_{2}\left\lceil x_{2}\right\rfloor ^{\frac{1}{2}}+\omega\left(x_{2},\, e_{2}\right)\\
\dot{e}_{1} & = & -l_{1}\left\lceil e_{1}\right\rfloor ^{\frac{2}{3}}+e_{2}\\
\dot{e}_{2} & = & -l_{2}\left\lceil e_{1}\right\rfloor ^{\frac{1}{3}}\\
\omega\left(x_{2},\, e_{2}\right) & = & k_{2}\left\lceil x_{2}\right\rfloor ^{\frac{1}{2}}-k_{2}\left\lceil x_{2}+e_{2}\right\rfloor ^{\frac{1}{2}}\,,
\end{array}\label{ec:CLError-1}
\end{equation}
is globally finite time stable.

We use the homogeneous and smooth LF
\begin{eqnarray*}
V\left(x,\, e\right)= & V_{1}\left(x\right)+\mu V_{2}\left(e\right) & \,,
\end{eqnarray*}
where $\mu>0$, 
\begin{eqnarray*}
V_{1}\left(x\right) & = & \gamma_{1}\left|x_{1}\right|^{\frac{5}{3}}+\frac{5}{2}\left(\frac{k_{1}}{k_{2}}\right)^{3}x_{1}x_{2}+\left|x_{2}\right|^{\frac{5}{2}}\,,
\end{eqnarray*}
\begin{eqnarray*}
V_{2}\left(e\right) & = & \left|\epsilon_{1}\right|^{\frac{5}{3}}+\gamma_{2}\left|e_{2}\right|^{\frac{5}{2}}\,,
\end{eqnarray*}
and $\epsilon_{1}=e_{1}-\frac{1}{l_{1}^{\frac{3}{2}}}\left\lceil e_{2}\right\rfloor ^{\frac{3}{2}}$.
$V_{2}$ is obviously p.d. for $\gamma_{2}>0$, while for $V_{1}$
it follows from Young's inequality (as in the previous subsection)
that it is positive definite selecting $\gamma_{1}$ sufficiently
large. The derivative of $V_{2}$ is
\begin{eqnarray*}
\dot{V}_{2}\left(e\right) & = & -\frac{5}{3}l_{1}\left\lceil e_{1}-\frac{1}{l_{1}^{\frac{3}{2}}}\left\lceil e_{2}\right\rfloor ^{\frac{3}{2}}\right\rfloor ^{\frac{2}{3}}\left(\left\lceil e_{1}\right\rfloor ^{\frac{2}{3}}-\frac{1}{l_{1}}e_{2}\right) 
 -\frac{5}{2}l_{2}\gamma_{2}\left\lceil e_{1}\right\rfloor ^{\frac{1}{3}}\left\lceil e_{2}\right\rfloor ^{\frac{3}{2}}\,.
\end{eqnarray*}
The first term is negative, except at the set ${\cal S}_{3}=\left\{ e_{1}=\frac{1}{l_{1}^{\frac{3}{2}}}\left\lceil e_{2}\right\rfloor ^{\frac{3}{2}}\right\} $,
on which the value of $\dot{V}_{2}$ is
\begin{eqnarray*}
\dot{V}_{2}\mid_{{\cal S}_{3}} & = & -\frac{5}{2}\frac{l_{2}}{l_{1}^{\frac{1}{2}}}\gamma_{2}\left|e_{2}\right|^{2}<0\,.
\end{eqnarray*}
And therefore, due to Lemma \ref{lemAndrieuH}, it is possible to
render $\dot{V}_{2}<0$ selecting $\gamma_{2}$ sufficiently small.
The derivative of $V_{1}$ along the trajectories of (\ref{ec:CLError-1})
is given by
\begin{eqnarray*}
\dot{V}_{1}\left(x\right) & = & \left(\frac{5}{3}\gamma_{1}\left\lceil x_{1}\right\rfloor ^{\frac{2}{3}}+\frac{5}{2}\left(\frac{k_{1}}{k_{2}}\right)^{3}x_{2}\right)x_{2}- 
 \frac{5}{2}k_{2}\left(\left(\frac{k_{1}}{k_{2}}\right)^{3}x_{1}+\left\lceil x_{2}\right\rfloor ^{\frac{3}{2}}\right)\left(\frac{k_{1}}{k_{2}}\left\lceil x_{1}\right\rfloor ^{\frac{1}{3}}+\left\lceil x_{2}\right\rfloor ^{\frac{1}{2}}\right)\\
 &  & +\frac{5}{2}k_{2}\left(\left(\frac{k_{1}}{k_{2}}\right)^{3}x_{1}+\left\lceil x_{2}\right\rfloor ^{\frac{3}{2}}\right)\omega\left(x_{2},\, e_{2}\right)\,.
\end{eqnarray*}
The second term is negative, except at the set $\left\{ x_{1}=-\left(\frac{k_{2}}{k_{1}}\right)^{3}\left\lceil x_{2}\right\rfloor ^{\frac{3}{2}}\right\} $,
at which the value of $\dot{V}_{1}$ is
\begin{eqnarray*}
\dot{V}_{1}\left(x_{1},\, x_{2}\right) & = & -\left(\frac{5}{3}\left(\frac{k_{2}}{k_{1}}\right)^{2}\gamma_{1}-\frac{5}{2}\left(\frac{k_{1}}{k_{2}}\right)^{3}\right)\left|x_{2}\right|^{2}\,,
\end{eqnarray*}
that is negative if $\gamma_{1}$ is sufficiently large, i.e.
\[
\gamma_{1}>\frac{3}{2}\left(\frac{k_{1}}{k_{2}}\right)^{5}\,.
\]
Thanks to Lemma \ref{lemAndrieuH} the first two terms in $\dot{V}$
can be made n.d. selecting $k_{2}$ sufficiently large. Note furthermore
that the function $\left\lceil x\right\rfloor ^{\frac{1}{2}}$ is
Hölder continuous, and therefore 
\[
\left|\omega\left(x_{2},\, e_{2}\right)\right|\leq ck_{2}\left|e_{2}\right|^{\frac{1}{2}}
\]
everywhere, for some $c>0$. We obtain therefore
\[
\dot{V}\leq-\alpha_{1}\left\Vert x\right\Vert _{\mathbf{r},\, p}^{4}+\alpha_{2}ck_{2}\left\Vert x\right\Vert _{\mathbf{r},\, p}^{3}\left\Vert e\right\Vert _{\mathbf{r},\, p}-\mu\alpha_{3}\left\Vert e\right\Vert _{\mathbf{r},\, p}^{4}\,,
\]
for some positive $\alpha_{i}$. Selecting $\mu$ sufficiently large
we obtain $\dot{V}<0$. Q.E.D.

We notice that system (\ref{ec:CLError-1}) is not insensitive to
the perturbation $\rho\left(t\right)$. However, as in the previous
case adding the discontinuous integral controller a Lipschitz continuous
perturbation can be completely compensated. The proof is similar to
that case.

\section{Simulation Example}

\label{Sec:sim} We illustrate the behavior of the proposed integral
controllers by some simulations. Consider the dynamics of a simple
pendulum without friction 
\[
\begin{array}{rcl}
\dot{x}_{1} & = & x_{2}\\
\dot{x}_{2} & = & -\frac{g}{l}\sin\left(x_{1}\right)+\frac{1}{ml^{2}}u+\rho\left(t\right)\,,
\end{array}
\]
where $x_{1}=\theta$ is the position angle, $x_{2}=\dot{\theta}$
is the angular velocity, $m$ is the mass of the bob, $g$ is the
gravity acceleration, $l$ is the length of the bob, the control $u$
is the torque applied to the pendulum, and the perturbation $\rho\left(t\right)=0.4\sin\left(t\right)$,
that can be interpreted also as the second derivative of a reference
signal (in this case the state $x$ corresponds to the tracking error).
For the simulations we have used the following parameter values $l=1\,[\text{m}]$,
$m=1.1\,[\text{Kg}]$, $g=9.815\,[m/s^{2}]$, and the initial conditions
$x_{1}\left(0\right)=2$, $x_{2}\left(0\right)=2$.

\begin{figure}[h]
\begin{centering}
\includegraphics[height=7cm]{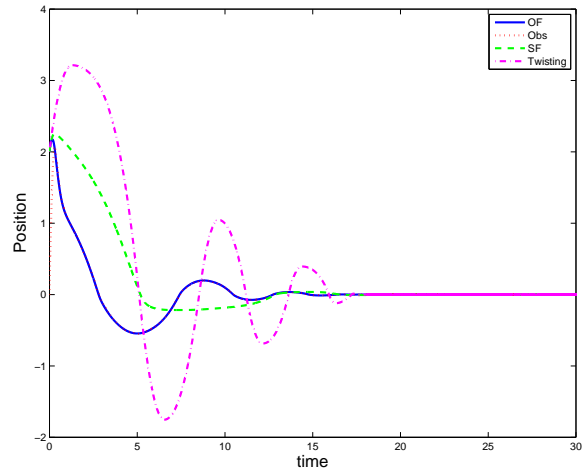} 
\caption{Behavior of $x_{1}$ with Twisting and the Discontinuous Integral Controller}
\end{centering}

\label{Pen1} 
\end{figure}

\begin{figure}[h]
\begin{centering}
\includegraphics[height=7cm]{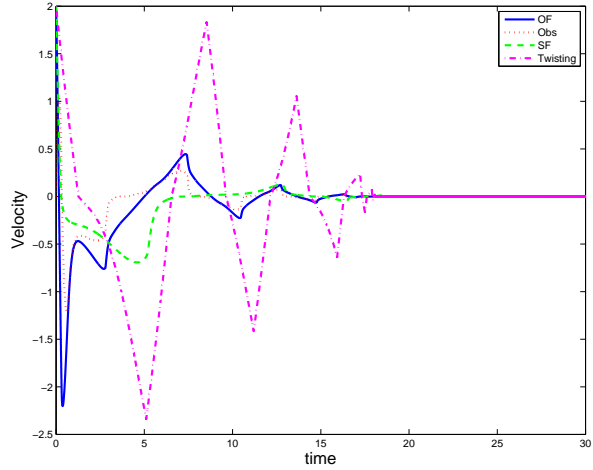} 
\caption{Behavior of  $x_{2}$ with Twisting and the Discontinuous Integral Controller}
\par\end{centering}

\label{Pen2} 
\end{figure}

\begin{figure}

\begin{centering}
\includegraphics[height=7cm]{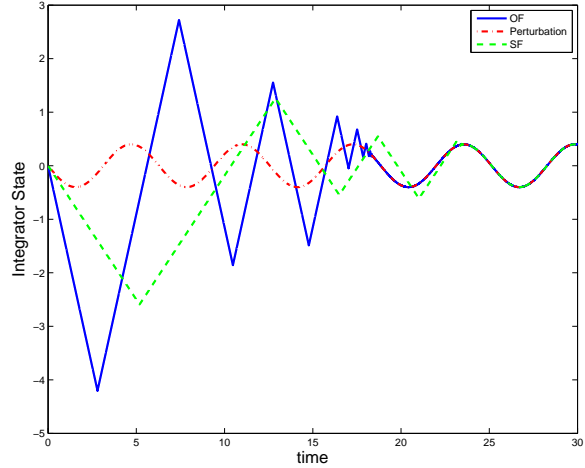}
\par\end{centering}
\caption{Behavior of the state of the discontinuous Integrator and the negative
value of the perturbation $-\rho\left(t\right)$. }
\label{Penz}

\end{figure}

\begin{figure}[h]
\begin{centering}
\includegraphics[height=7cm]{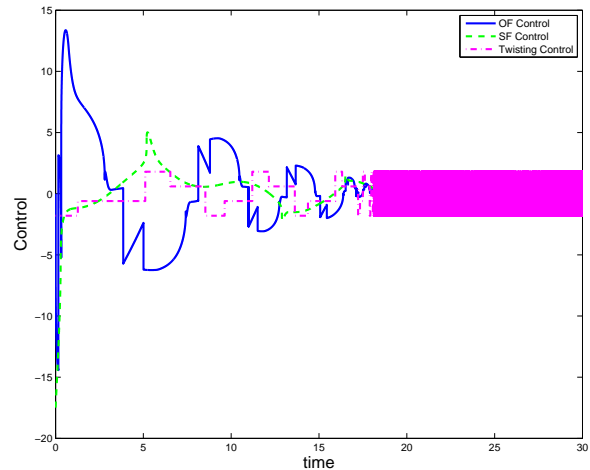}

\par\end{centering}

\caption{Control signal for OF and SF controllers with discontinuous Integral
action and for the Twisting controller.}

\label{PenU} 
\end{figure}

We have implemented three controllers:
\begin{itemize}
\item A State Feedback (SF) controller with discontinuous integral term,
as given by (\ref{ec:CID}), with gains $k_{1}=2$, $k_{2}=5$, $k_{3}=0.5$,
$k_{4}=0$, and initial value of the integrator $z\left(0\right)=0$.
\item An Output Feedback (OF) controller with discontinuous integral term,
as given by (\ref{ec:CID-Obs}), with controller gains $k_{1}=2\lambda^{\frac{2}{3}}$,
$k_{2}=5\lambda^{\frac{1}{2}}$, $k_{3}=0.5\lambda$, $k_{4}=0$,
$\lambda=3$, observer gains $l_{1}=2L$, $l_{2}=1.1L^{2}$, $L=4$,
observer initial conditions $\hat{x}_{1}\left(0\right)=0$, $\hat{x}_{2}\left(0\right)=0$,
and initial value of the integrator $z\left(0\right)=0$.
\item A Twisting controller \cite{Lev93,FriLev02,Lev03}, given by $u=-k_{1}\left\lceil x_{1}\right\rfloor ^{0}-k_{2}\left\lceil x_{2}\right\rfloor ^{0}$,
with gains $k_{1}=1.2$, $k_{2}=0.6$.
\end{itemize}
The simulations for the three controllers are presented in Figures
\ref{Pen1}-\ref{PenU}. In Figure \ref{Pen1} the evolution of the
position is presented and also the evolution of the estimated position
given by the observer for the OF, which converges very fast. All controllers
are able to bring the position to zero in finite time.

Figure \ref{Pen2} presents the time evolution of the velocity and
its estimation by the observer for the OF, which converges in finite
time around the time $15$. We see also the typical zig-zag behavior
for the Twisting controller. All controllers are able to bring the
velocity to rest in finite time.

In Figure \ref{Penz} the integrator state is presented for both controllers
OF and SF, and the (negative) value of the perturbation ($-\rho\left(t\right)$).
We note the zig-zag behavior of the integral controller, due precisely
to its discontinuous character. We appreciate also that the integrator
signal reconstructs after a finite time the (negative value of the)
perturbation, and this is the reason for it to be able to fully compensate
its action on the plant.

Finally, Figure \ref{PenU} presents the control signal $u$ for the
three controllers. We see that, while the OF and the SF controllers
with discontinuous Integral action provide a continuous control signal,
the Twisting controller provides a switching (discontinuous) control
signal, with an extremely high frequency when the equilibrium has
been reached, which corresponds to the (undesirable) chatttering phenomenon.

\section{Conclusion}

\label{sec:Conclusion}We present in this paper a Discontinuous Integral
Controller, which shares the properties of the classical PID control
and the HOSM controllers: Similar to HOSM it is able to fully compensate
a Lispchitz perturbation or to track an (unknown) time varying reference
with bounded second derivative, it has high precision due to the homogeneity
properties, and it stabilizes the origin globally and in finite time.
Similar to the PID control it has a continuous control signal. In
order to achieve an Output Feedback scheme we introduce a finite time
converging observer. The stability proofs are performed with a novel
Lyapunov method. It is possible to extend this idea to systems with
higher relative degree, and this will be done in future work.

\section*{Acknowledgement}

The authors thank the financial support from PAPIIT (UNAM), project
IN113614; Proyecto de Cooperación II-FI, Project IISGBAS-122-2014;
and CONACyT, project 241171. 

\bibliographystyle{amca}

\begin{thebibliography}{10}
\bibitem{Kha02}Khalil, H. K. (2002). {\em Nonlinear Systems}.
3rd Ed., Prentice Hall.

\bibitem{Kok92}Kokotovic, V. P. (1992). The joy of feedback: nonlinear
and adaptive. {\em IEEE Control Systems}. 7--17.

\bibitem{ZenEls00}Zenieht, S. y A. L. Elshafe (2000). Robust control
of uncertain nonlinear Mechanical systems using high gain observer.
{\em Proceedings of the American Control Conference}. Chicago.
pp. 3621--3625.

\bibitem{Mor12}Moreno, Jaime A. (2012). Lyapunov function for Levant's
second order differentiator, {\em Proceedings of the 51st IEEE Conference
on Decision and Control}.Maui, Hawai, Dec. 10-13,2012.

\bibitem{Utk09}Utkin, V. (2009).{\em Sliding Mode Control in Electro-Mechanical
Systems}. CRC Press. Second edition. Automatic and Control Enineering.

\bibitem{Utk92}Utkin, V. (1992). {\em Sliding Modes in Control
and Optimization}. Springer-Verlag.

\bibitem{FriLev02}Fridman, L. y A. Levant (2002).{\em Sliding Mode
in Control in Engineering}. Marcel Dekker, Inc. High Order Sliding
Modes.

\bibitem{BacRos05}Baccioti, A. and L. Rosier (2005). {\em Liapunov
functions and stability in control theory}. Ed 2$^{nd}$. New York.
Springer-Verlag.

\bibitem{Fil98}Filippov, A. F. (1998).{\em Differential equations
with discontinuous right hand side}. Kluwer. Dordrecht. The Netherlands.

\bibitem{HarLit51}Hardy, G. H., J. E. Littlewood y G. Polya (1951).
{\em Inequalities}. London. Cambridge University Press.

\bibitem{Lev07}Levant, A. (2007). Principles of 2-sliding mode design.
{\em Automatica}. \textbf{43}, 576--586.

\bibitem{DavFri06}Davila, J., L. Fridman; A. Poznyak (2006). Observation
and identification of mechanical systems via second oreder sliding
modes. {\em Int. J. of Control}. \textbf{79}, 1251--1262.

\bibitem{Lev05}Levant, A. (2005). Homogeneity approach to higher-order
sliding mode design. {\em Automatica}. \textbf{34}, 576--586.

\bibitem{Lev03}Levant A. Higher order sliding modes: differentiation
and output feedback control. \emph\{Int. J. of Control\};
nov 2003; \textbf\{76\} (9-10):924--941.

\bibitem{MorOso12}Moreno, Jaime A.; M. Osorio (2012). Strict Lyapunov
functions for the super-twisting algorithm. {\em IEEE Transactions
on Automatic Control}. \textbf{54}, 1035--1040.

\bibitem{MorOso08}Moreno, Jaime A. Osorio, M. (2008). A Lyapunov
approach to second-order sliding mode controllers and observers. {\em
Proceedings of the 47th IEEE Conference on Decision and Control}.
Cancún, México. Dec. 9--12.

\bibitem{Lev93}Levant, A. (1993). Sliding order and sliding accuracy
in sliding mode control. {\em International Journal of Control}.
\textbf{6}, 1247--1263.

\bibitem{Hes66}Hestenes, M. R. Calculus of variations and optimal
control theory. John Wiley \& Sons, New York, 1966. 

\bibitem{AndPra08}V. Andrieu, L. Praly, A. Astolfi. Homogeneous approximation,
recursive observer design and output feedback. SIAM J. Con. Opt.,
2008, \textbf{47}(4):1814--1850. 

\bibitem{ZamMor13}Zamora, Cesar; Moreno, J. A.; Kamal, Shyam. Control
Integral Discontinuo Para Sistemas Mecánicos. Congreso Nacional de
Control Automático 2013. Asociación de México de Control Automático
(AMCA). Ensenada, Baja California, 16-18 Octubre 2013. Pp. 11--16.
http://eventos.cicese.mx/amca2013/52.html
\end{thebibliography}

\end{document}